\documentclass[conference]{IEEEtran}

\usepackage[latin1]{inputenc}

\usepackage{amsmath,bbm,comment,bbm}
\usepackage{graphicx}

\def\N{\mathbbm{N}}

\def\E{\mathbbm{E}}

\newtheorem{lemma}{Lemma}
\newtheorem{proposition}{Proposition}
\newcommand{\n}{{\rm \bf n}}

\begin{document}

\title{Impatience in mobile networks and its application to data pricing}
\author{

\IEEEauthorblockN{Fabrice Guillemin, Salah Eddine Elayoubi} 
\IEEEauthorblockA{Orange Labs,       France \\        First-Name.Name@orange.com}
\and
\IEEEauthorblockN{Philippe Robert, Christine Fricker, Bruno Sericola} 
\IEEEauthorblockA{INRIA, France \\  First-Name.Name@inria.fr} \\

}

\maketitle
\begin{abstract}
We consider in this paper an import Quality of Experience (QoE) indicator in mobile networks that is reneging of users due to impatience. We specifically consider a cell under heavy load conditions and compute the reneging probability by using a fluid limit analysis. By solving the fixed point equation, we obtain a new QoE perturbation metric quantifying the impact of reneging on the performance of the system. This metric is then used to devise a new pricing scheme accounting of reneging. We specifically propose several flavors of this pricing around the idea of having a flat rate for accessing the network and an elastic price related to the level of QoE perturbation induced by  communications.
\end{abstract}

\section{Introduction}

An important aspect of the quality of experience (QoE)  in cellular networks, which is in general a complex mixture of many parameters,  is the reneging by users. Reneging results from impatience when users feel that their  communications last for an excessive amount of time. Such a phenomenon is negative for both the user and the network  as radio resources are uselessly consumed by impatient users before they abort their communications.

Impatience has been the object of several works dealing with fixed networks. In \cite{baccelli},  a new version of  Erlang formula has been derived and is applicable for the case of streaming-like flows, where the service duration is independent of the quantity of resources obtained by the user. This is a key difference with the case of data traffic considered in this paper, where service duration is proportional to the quantity of resources obtained by the user.  In \cite{yang01}, the authors model impatience using the deterministic service curves approach  and used simulations to quantify its impact on the system performance for several bandwidth sharing disciplines.  

In \cite{bonald}, the authors analyze data traffic at the flow level and consider impatience of users in the overload regime, when the mean arrival intensity is larger than the mean service rate. Along the same line of investigations, we develop in this paper a fluid flow analysis of impatience in cellular networks. We notably establish  a fixed point formulation for the computation of  the reneging probability for users sharing bandwidth of a cell under overload conditions. These reneging probabilities are then used to introduce a new metric, namely QoE perturbation, expressing how much a particular flow impacts the reneging probability in the system. We then use this QoE perturbation metric to design of a new pricing framework. 

The main contributions of this paper are as follows:
\begin{itemize}
\item We develop a tractable analytic model for impatience of users in mobile networks, from both network  and user perspectives (steady-state probabilities and probability of reneging);
\item We develop a QoE perturbation model based on the impact of a new communication on the reneging probability of users in the cell;
\item We discuss how this QoS perturbation scheme can be exploited for  pricing.
\end{itemize}

 In Section~\ref{model}, we describe the system under consideration and introduce impatience from a network point of view.  On the basis of this analysis, we devise in Section~\ref{numeric} new pricing schemes accounting of impatience. Concluding remarks are presented in Section~\ref{conclusion}.

\section{System model with impatience}
\label{model}
\subsection{Basic model with no impatience}
\begin{figure}[ht]
\centering
\includegraphics[width=8cm]{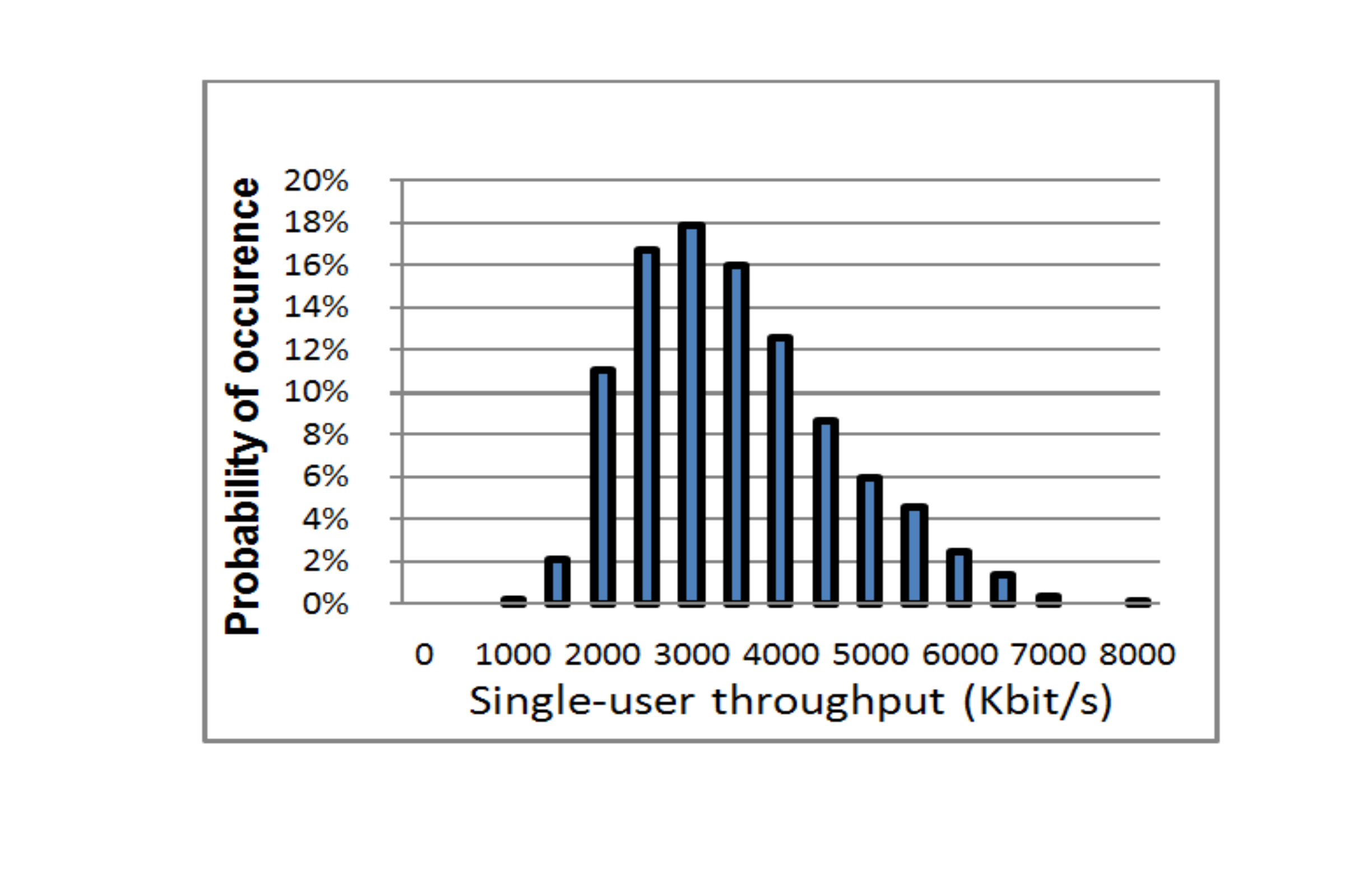}
 \caption{Probability distribution function of throughput obtained from drive test measurements in an HSDPA network in a large European city (category 14 devices with LMMSE receivers).}
\label{debitdistrib}
\end{figure}

We focus on one cell of a cellular networks and consider a general model, applicable to 3G and 4G networks, where resources (time slots in HSDPA and Resource Blocks in LTE) are equally divided among users. Because of propagation and interference, the capacity at the edge of a cell is lower than that at its center, as illustrated in Figure \ref{debitdistrib} where the throughput distribution obtained from drive test measurements is represented. This corresponds to a multi-class system where users at different positions belong to different classes.  Let $K$ be the number of possible classes of radio conditions, where class $k\in\{1,\ldots, K\}$ is characterized by a throughput $c_k$ ($c_1>c_2> \ldots >c_{K}$) and has a weight in the total traffic demand equal to $p_k$ such that $\sum_{k=1}^K p_k=1$.

We assume that users present in the cell download data, thus giving rise to data flows (typically TCP connections).  In the following, we assume that flows appear according to a Poisson process with rate $\lambda$. The cell capacity is shared among the various data flows according to the processor sharing discipline. If a data flow originates in ring $k$, the instantaneous service rate for this flow is equal to $c_k/n$ if there are globally $n$ active flows.

Let us assume that the volume $\sigma$ of data per flow has a general 
distribution with mean $\E(\sigma)$. Given the heterogeneity of the bit rates 
available to users according to their location, we are led to consider a 
multiclass processor sharing system, where a customers of class $k$ has a 
requested service time equal to $T_k = \sigma/c_k$. In the following we 
define $\mu_k$ by $\mu_k = 1/\E(T_k) = c_k/\E(\sigma)$. 

The steady state probability of the multiclass processor sharing queue is known in the  literature to have a product form \cite{Ross}. Specifically, the steady state probability that there are $n_k$ flows in progress in class $k$ for $k=1,\ldots,K$ is 
$$
P(\mathbf{n}) = \frac{1}{1-\rho} 
\frac{|\mathbf{n}|!}{\mathbf{n}!}\prod_{k=1}^K \rho_k^{n_k},
$$
where $\mathbf{n} =(n_1,\ldots,n_K)$, $\mathbf{n}!=n_1!\ldots n_K!$, 
$|\mathbf{n}|=n_1+\ldots+n_K$, $\rho_k=\lambda p_k/\mu_k$ and $\rho=\sum_{k=1}^K \rho_k$. 
The stability condition of the system reads $\rho <1$.

In the following, we consider for $n \geq 0$ the set $\mathcal{S}_n$ composed of those K integer tuples $(n_1,\ldots,n_K)\in \N^K$ such that $n_1+\ldots +n_K=n$. The cardinal of $\mathcal{S}_n$ is   $|\mathcal{S}_n| = \left( \begin{array}{c} n+K-1 \\ n\end{array}  \right)$.

\subsection{Modeling impatience}

We suppose that users may abort their download if the transmission of 
data takes too much time. We specifically assume that users renege at the 
expiration of some timer with duration $\tau$ independent of the 
transmission time. We further assume that the random variable $\tau$ 
is exponential with mean $1/\mu_0$ and  that $\mu_0 < \mu_K$. This assumption means that a user is ready to wait more than the time needed for the completion of the download if he were alone in the cell in the worst radio conditions.

Let $n_k(t)$ denote the number of customers (i.e., the number of active flows) 
of class $k$ in the system at time $t$. Assuming that the amount $\sigma$ of data to 
transmit  is exponentially distributed, the process 
$((n_1(t),\ldots,n_{K}(t)))$ is a Markov process. 

Let $q(\mathbf{n},\mathbf{m})$, where $\mathbf{n},\mathbf{m} \in \N^K$, 
denote the transition rate from state $\mathbf{n}$ to state $\mathbf{m}$. 
The non-null transition rates are 
$$
q(\mathbf{n},\mathbf{n}+\mathbf{e}_k) = \lambda p_k, \quad \mbox{and} \quad q(\mathbf{n},\mathbf{n}-\mathbf{e}_k)  = \frac{n_k\mu_k}{|\mathbf{n}|} +n_k\mu_0,
$$
for $k = 1,\ldots,K$, where $\mathbf{e}_k$ is the row vector with all components equal to 0 except the $k$th one equal to 1.
Furthermore, we set $q(\mathbf{n},\mathbf{n})= -\sum_{\mathbf{m}\neq \mathbf{n}} q(\mathbf{n},\mathbf{m})$.

The number of customers in the system is less than the number of customers in an $M/M/\infty$ with arrival $\lambda$ and service rate $\mu_0$. This implies that the system under consideration is stable even if $\rho >1$. There exists a  unique  invariant probability distribution given by the row vector $(\pi(\n), \n \in \N^K)$ which satisfies
\begin{equation}
\label{pistat}
\pi Q = 0 \quad \mbox{and} \quad \sum_{\n\in \N^K} \pi(\n)=1.
\end{equation}

\subsection{Probability of reneging}
\label{renegsection}

Let $P_k(\mathbf{n})$ be the probability that a customer of class $k$ reneges while there are $n_\ell$ customers of class $\ell = 1,\ldots,K$ in the system upon its arrival so that $\mathbf{n}=(n_1, \ldots, n_K)$. By using the memoryless property of the exponential distribution, we can easily prove the following result.

\begin{lemma}
The probabilities $P_k(\n)$ for $\n \in \N^K$ satisfy the recurrence relations
\begin{multline}
\label{recPk}
P_k(\mathbf{n}) =\frac{\mu_0}{\Lambda_k(\mathbf{n})}+\sum_{\ell=1}^K \frac{\lambda p_\ell}{\Lambda_k(\mathbf{n})} P_k(\mathbf{n}+\mathbf{e}_\ell) \\ + \sum_{\ell=1}^K 
\frac{1}{\Lambda_k(\mathbf{n})}\left(\frac{n_\ell\mu_\ell}{|\mathbf{n}|+1}+n_\ell\mu_0 \right) P_k(\mathbf{n}-\mathbf{e}_\ell) ,
\end{multline}
where 
$$
\Lambda_k(\mathbf{n}) = \lambda+\sum_{\ell=1}^K\frac{(n_\ell+\delta_{k,\ell})\mu_\ell}{|\mathbf{n}|+1} + \left(|\mathbf{n}|+1\right)\mu_0
$$
with $\delta_{k,\ell}$ denoting the Kronecker symbol.
\end{lemma}

The recurrence relation~\eqref{recPk} can be rewritten in matrix form as
\begin{equation}
\label{renegmat}
(\mathbbm{I}-M_k)\mathbf{P}_k=\mathbf{u}_k,
\end{equation}
where $\mathbbm{I}$ is the identity matrix, $\mathbf{P}_k$ (resp. $\mathbf{u}_k$) is the column vector with components $P_k(\mathbf{n})$ (resp. $\mu_0/\Lambda_k(\mathbf{n})$),  $\mathbf{n} \in \N^K$, and the matrix $M_k$ is given by
\begin{equation}
\label{defMk}
M_k = \left( \begin{array}{ccccc}
0 & A_{k,0} & & & \\
B_{k,1} & 0 & A_{k,1} & & \\
& B_{k,2} & 0 & A_{k,2} & \\
& & . & . &.
\end{array} \right)
\end{equation}
with  matrices $A_{k,n}$ and $B_{k,n}$ being defined as follows:
\begin{itemize}
\item The non-null entries of  matrix $A_{k,n}$ are defined for $n \geq 0$ and $\mathbf{n}=(n_1, \ldots, n_K)\in \mathcal{S}_n$ by for $j=1,\ldots,K$
$$
a_{k,n}(\mathbf{n},\mathbf{n}+\mathbf{e}_j) = \frac{\lambda p_j}{\Lambda_k(\mathbf{n})};
$$
\item The non-null entries of  matrix $B_{k,n}$ are defined for $n \geq 0$ and $\mathbf{n}=(n_1, \ldots, n_K)\in \mathcal{S}_n$ by for $j=1,\ldots,K$
$$
b_{k,n}(\mathbf{n},\mathbf{n}-\mathbf{e}_j) = \frac{1}{\Lambda_k(\mathbf{n})}
\left(\frac{n_j\mu_j}{|\mathbf{n}|+1} + n_j \mu_0\right).
$$
\end{itemize}

In spite of the fact that the norm of the matrix $M_k$ is 
$$
\|M_k\| =\sup_{\mathbf{n}\in \N^K}\sum_{\mathbf{m} \in \N^K} |M_k(\mathbf{n},\mathbf{m})|=1,
$$
we can show the following result by using the fact that  $0\leq P_k(\mathbf{n}) \leq 1$ for all $\mathbf{n} \in  \N^K$.

\begin{proposition}
The vector $\mathbf{P}_k$ is given by
\begin{equation}
\label{Pkexp}
\mathbf{P}_k = \sum_{r=0}^\infty (M_k)^r \mathbf{u}_k,
\end{equation}
where $\mathbf{u}_k$ is introduced in Equation~\eqref{renegmat}. 
\end{proposition}

The reneging probability $\mathcal{P}_k$ that a class $k$ customer reneges is eventually given by
\begin{equation}
\label{renegprobk}
\mathcal{P}_k = \pi . \mathbf{u}_k,
\end{equation}
where $\pi$ is the row vector satisfying Equation~\eqref{pistat}.

Since Equation~\eqref{renegmat} may be very difficult to solve since we have to deal with the infinite matrix $M_k$, we develop in the next section an approximating model when the load $\rho >1$. For impatience, this case is the most interesting as many customers may renege; impatience may have a negligible impact when $\rho<1$. Note that the case $\rho \sim 1$ requires a more detailed analysis.

\section{QoE-perturbation based pricing scheme}
\label{numeric}
We now move to a practical application of the model developed in this paper that consists in developing a QoE perturbation metric and its usage in the design of a pricing scheme.
\subsection{Classical pricing schemes and their shortcomings}

Pricing of data communications in cellular networks is usually based on capped offers. The user can download up to a certain threshold and is then throttled independently of the state of the network when the volume of transmitted data  exceeds the prescribed threshold. The price that the user has to pay precisely depends on this threshold. This is a major difference with fixed networks, where pricing is based on flat rates. 

With the rapid development of mobile Internet and the deployment of  4G technology, capped offers will however not be well suited  to the new usage of wireless communications, which offer a performance comparable to that in fixed networks. This is all  the more true as with the widespread of smart phones and tablets, users do not care so much of the way they are accessing the Internet, via cellular or WiFi/fixed access. The major drawback of capped offers is that they do not account of network state and they require the deployment of a monitoring infrastructure, which can be very costly and leads to the concentration of traffic in gateways, namely  Packet Data Network (PDN) gateways in the 4G architecture.

In this paper, we investigate new ways of pricing data  communications by going beyond capped offers as used in today's cellular data networks or pricing based on usage as in classical telephony networks. We instead exploit the fact that the capacity of a cell is shared between all active users thanks to the scheduling algorithm implemented in the base station. A first idea in this context would be to charge the user according to the average bit rate he obtains for his communications (bit rate  pricing). This is however unfair in several respects. For instance, the achieved bit rate depends on the location of the user in the cell, which is a parameter not completely under control by the user. Moreover, the higher the bit rate for a communication, the shorter is the communication and hence the smaller is the impact of this user on other users. 

An alternative approach would be to introduce a social  welfare function for the various customers present in a cell and use Vickrey-Clarke-Groves  pricing model \cite{VCG,yang}. As shown in \cite{Birmiwal},  when customers share a common transmission capacity $c$  under the fair share policy (i.e., the welfare function for a customer is proportional to $\log (x)$ where $x$ is the achieved bit rate) then the price to pay is an increasing and concave function of the achieved rate $x$. This scheme has hence the same shortcomings as bit rate pricing. Other pricing schemes based on rate allocation are also investigated in \cite{Birmiwal}.

Congestion could also be used for pricing as in Conex scheme that supposes that traffic sources (using for instance TCP) adapt to congestion signals \cite{conex}. However, in today's networks, TCP is more and more bursty and may create backlog even if the network is not congested (in terms of average offered traffic). In addition to this, the bottleneck in wireless networks is often the radio access so that congestion based on buffer levels as in Conex is  difficult to assess because of fluctuations of the available bandwidth. Moreover, customers in bad radio conditions may create backlog in the base station but this is not a sign of congestion; packets accumulate only because the bit rate achievable by the customer is small. 

We propose in this section a new pricing scheme based on the notion of social cost of a user, or the QoE perturbation induced by the arrival of this user to the cell. We begin by describing the metric of QoE perturbation before proposing the pricing scheme.

\subsection{QoE perturbation metric}

We begin by introducing the QoE perturbation measure caused by a communication. 
QoE perturbation is related to the impact of the presence of a given flow on the QoE of other users. Recall that $P_j({\bf n})$ is the probability of reneging for a class $j$ customer when there are ${\bf n}$ customers in the system. The extra perturbation introduced by a class $k$ customer when the system is in state ${\bf n}$ can be computed by:
\begin{equation}
\Gamma_k({\bf n})=\sum_{j=1}^K p_j (P_j({\bf n}+{\bf e_k})-P_j({\bf n})).
\end{equation}
This means that the QoE perturbation for class $k$ users is the average additional impatience observed by users when there is an additional user of class $k$ in the system. Note that this QoE perturbation depends on the state of the system, leading to a pricing scheme that depends on the instantaneous state of the system. It is thus preferable to define an average QoE perturbation metric as follows:
\begin{equation}
\hat{\Gamma}_k(\lambda_1,...,\lambda_K)=\sum_{{\bf n}} \Gamma_k({\bf n}) \pi({\bf n}).
\end{equation}

Note that this metric may be complex to compute, as it involves the computation of the reneging probabilities. We then make use of the approximation of the previous section and define the QoE perturbation function for class $k$ users as follows:
\begin{equation}
\tilde{\Gamma}_k(\lambda_1,...,\lambda_K)=\frac{\partial\tilde{R}}{\partial \lambda_k},
\end{equation}
where the global impatience rate is defined as follows:
\begin{equation}
\tilde{R}(\lambda_1,...,\lambda_K)=\sum_{j=1}^K \lambda_j \widetilde{\mathcal{P}}_j(\lambda_1,...,\lambda_K)= S(\lambda_1,...,\lambda_K).
\end{equation}

In the general case of $K$ radio conditions, it is sufficient to derive the fixed point equation (\ref{pointfixe}), see~\cite{EFGRS3},
\begin{equation}
\label{pointfixe}
\sum_{k=1}^K\frac{\lambda_k}{\mu_k+S}=1.
\end{equation}

in order to obtain the QoE perturbation metric:
\begin{equation}\label{QoE_perturb}
\tilde{\Gamma}_k(\lambda_1,...,\lambda_K)=\frac{\partial S}{\partial \lambda_k}=\left((\mu_k+S)\sum_{j=1}^K\frac{\lambda_j}{(\mu_j+S)^2}\right)^{-1}
\end{equation}

\begin{figure}[hbtp]
\begin{center}
\scalebox{0.45}{\includegraphics{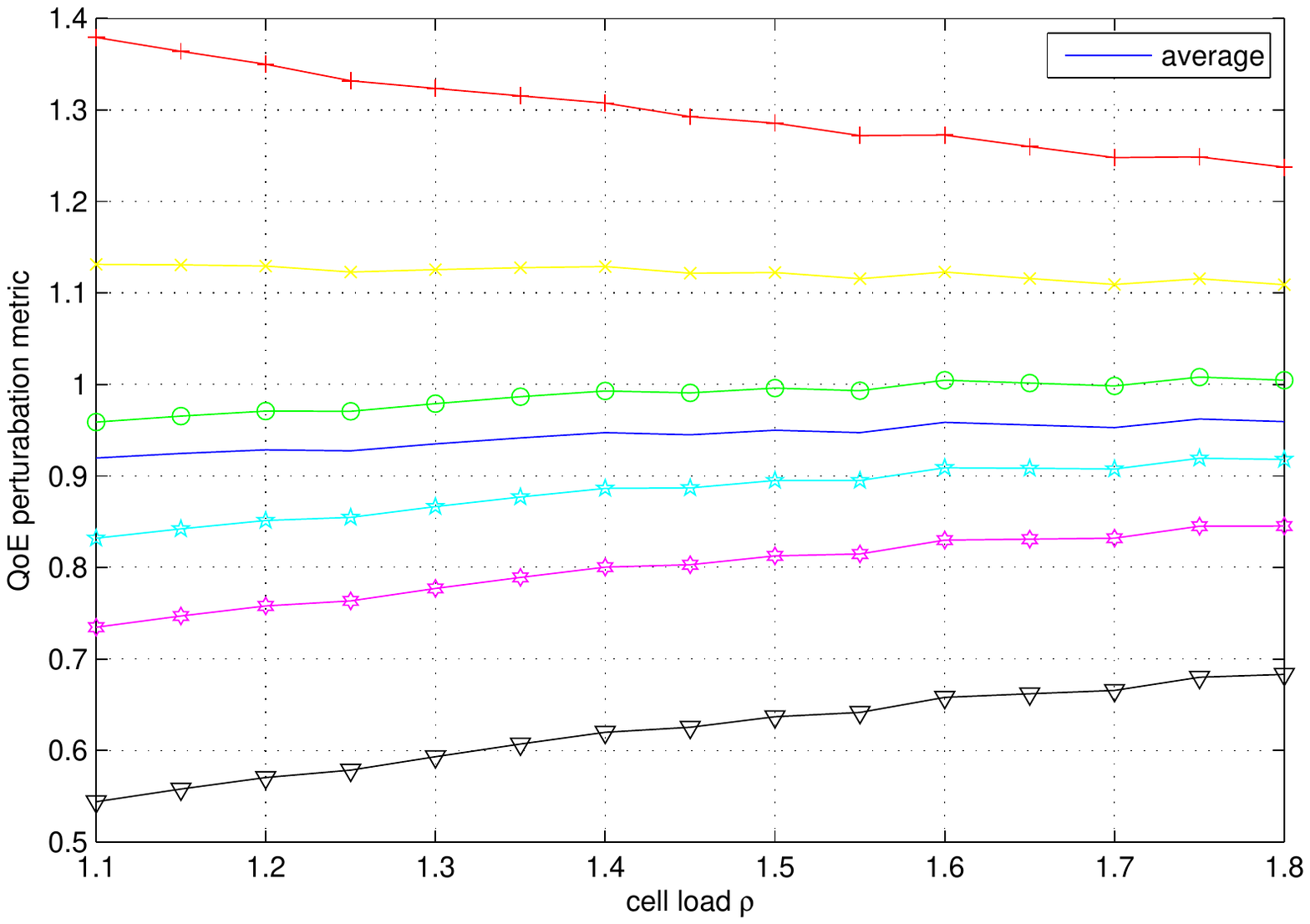}}
\end{center}
\caption{QoE perturbation metric for users belonging to different classes.\label{price_general}}
\end{figure}

Figure \ref{price_general} illustrates the QoE perturbation caused by the different classes of radio conditions (taken from figure \ref{debitdistrib}). The following properties can be observed:
\begin{itemize}
\item QoE perturbation is larger for cell edge users than for cell center users. Cell edge users contribute more to the reneging than cell center users.
\item QoE perturbation is almost a flat function of the cell load, even if the impact of cell edge users reduces at large loads (when the impact of radio conditions reduces compared to the impact of increasing number of active users). This is an important property as the operator can set a fixed approximate QoE perturbation metric function of the radio condition and independent of the traffic load.
\item QoE perturbation does not depend on the impatience rate (this is not observed from the figure but directly from the mathematical formula (\ref{QoE_perturb}) as $S$ is independent of $\mu_0$). This is an important property as the operator does not have to estimate the impatience rate, difficult to be assessed from field measurements, but only the distribution of radio conditions.
\end{itemize}

\subsection{Proposed pricing scheme}
Based on this QoE perturbation metric and for a convergence of fixed and cellular networks in terms of pricing, we propose a pricing scheme comprising two components:
\begin{itemize}
\item A flat rate to enable a customer to attach to the network (authentication, roaming, etc.);
\item An elastic component reflecting the ``social'' cost of a user, i.e. depending on the QoE perturbation metric defined above.
\end{itemize} 

Different possibilities exist for the elastic component. A first possibility is to define a per-Mbyte price for the different radio conditions, proportional to the QoE perturbation metric. A second possibility, more inline with the current capped offers and the Conex strategy, would be to define a \lq\lq{}congestion right limit\rq\rq{} that is decremented proportionally to the QoE perturbation associated with the radio condition. A pricing notification mechanism could also be implemented, where the users receive, depending on their position and the network state, their price/congestion information.

\section{Conclusion}
\label{conclusion}

By using a fluid limit approximation, we have studied in this paper the reneging probability of customers sharing the radio resources in mobile networks. We show the reneging probability can be easily derived using a simple fixed point equation and derive a QoE perturbation metric that corresponds to the impact a particular communication has on the QoE of other users. We then introduced new pricing policies that comprise a fixed component (flat rate as in fixed networks) and an elastic one depending on the QoE degradation metric.


\end{document}